\documentclass{knawproc}

\input{epsf}

\begin{document}

% The ``opening'' environment takes care of title, author and headlines
\begin{opening}

\title{Semianalytic modelling of the formation and evolution of galaxies}

\author{C.M. Baugh$^1$, C.G. Lacey$^2$, S. Cole$^1$, \& C.S. Frenk$^1$}
\addresses{
  1. Department of Physics, South Road, Durham University, DH1 3LE, UK\\
  2. Theoretical Astrophysics Center, Juliane Maries Vej 30, DK-2100 
     Copenhagen \O, Denmark \\
}
\runningtitle{Semianalytic galaxy formation}
\runningauthor{Baugh et al.}

\end{opening}

%%%%%%%%%%%%%%%%%%%%%%%%%%%%%%%%%%%%%%%%%%%%%%%%%%%%%%%%%%%%%%%%%%%%%%%%%%%%%%%

\begin{abstract}
The high redshift observations of galaxies now becoming 
available from the Hubble Space Telescope and from large ground 
based telescopes are opening fresh windows on galaxy formation.
Semianalytic models of galaxy formation provide us with a powerful 
tool to interpret and understand these exciting new data.  
In this review, we explain the philosophy behind this class of model 
and outline some of their remarkable successes, focussing our attention 
on the formation of elliptical galaxies and on the properties of galaxies at 
high redshift.
Now that the recent discovery of star forming galaxies at $z \sim 4$ 
has made possible the construction of the cosmic star formation history, 
which is in good agreement with our model predictions, it appears that 
a coherent picture of galaxy formation is beginning to emerge.
\end{abstract}

%%%%%%%%%%%%%%%%%%%%%%%%%%%%%%%%%%%%%%%%%%%%%%%%%%%%%%%%%%%%%%%%%%%%%%%%%%%%%%%

\section{Introduction}

The fundamental questions of `How and when do galaxies form?' and 
`What are the major influences that determine their appearance?' 
are still unresolved. 
However, with the development of powerful theoretical techniques and the 
increasing availability of high redshift observations, impressive 
progress is being made towards changing this situation.

In the traditional approach to modelling galaxy evolution, pioneered 
by Tinsley (1980), a set of local galaxy templates are combined 
in the locally observed number densities to make predictions of the faint 
galaxy counts and redshift distributions. 
Simple {\it ad hoc} parameterisations can be made to describe the evolution of 
the luminosity and number density of galaxies in order to improve 
the fit to the deepest observed counts. In this retrospective approach, 
the formation epoch of galaxies is placed at some arbitrary high redshift. 

Since the start of the 1980's however, our understanding of the growth 
of structure in the universe has increased enormously. 
In the currently favoured cosmologies, the universe is gravitational 
dominated by some form of dark matter. Dark matter halos grow 
hierarchically through mergers and accretion. 
The formation of structure through the gravitational 
amplification of small, primordial density fluctations is now well 
understood, mainly as a result of large numerical N-body simulations 
of this process ({\it e.g.} for a recent example see Jenkins et al 1998). 
The growth of dark matter halos can be equally well described 
analytically, at least in a statistical sense, via the theory 
developed by Press \& Schechter (1974) and its extensions 
(Bond et al 1991, Bower 1991), as demonstrated by comparison with 
N-body simulations by Lacey \& Cole (1994). 

The analytical description of hierarchical clustering can be used 
to construct Monte-Carlo realisations of the complete merger 
history of dark matter halos. 
The merger history of a halo is then combined with a set of simple 
rules that encapsulate our present understanding of 
the processes involved in galaxy formation: 

\begin{itemize}

\item The cooling and condensation of gas within dark matter 
halos (Rees \& Ostriker 1977, Silk 1977, Binney 1977; White \& Rees 1978).

\item Star formation from the reservoir of gas that cools during the 
halo lifetime. 

\item Feedback process, such as supernovae and stellar winds, 
that regulate the star formation. This is necessary in hierarchical 
models to prevent all the gas from cooling and forming stars in small, 
dense objects at high redshift in which cooling is very 
efficient (White \& Rees 1978, Cole 1991, White \& Frenk 1991).

\item Mergers of galaxies -- galaxies can coalesce on a much longer 
timescale than their host dark matter halos.

\item The conversion of star formation histories into spectra and broad band 
luminosities using stellar population models ({\it e.g.} Bruzual \& 
Charlot 1993).

\end{itemize}

The result is a physically motivated, semianalytic model that is driven  
by structure formation in the universe. The inputs of the traditional 
models, the luminosity function and morphological mix of galaxies, 
are actually {\it predicted} by semianalytic models.
Contrary to first expectations, surprisingly few parameters are needed to 
specify the model, once a cosmology has been adopted. These parameters   
are set by reference to a subset of local observations ({\it e.g.} 
matching to $L_{*}$ of the B-band luminosity function in the model of 
Cole et al 1994, or putting a galaxy in a Milky Way sized dark matter halo 
on the Tully-Fisher relation in the models of Kauffman, White \& 
Guiderdoni 1993). 
The remaining output, number counts, colours, redshifts distributions 
etc., is then predictive. 

It is important to realise that the models do not set out to make 
a `fit' to an observable such as the local luminosity function in the 
usual sense -- rather we attempt to choose the physical parameters to 
achieve the best {\it match} or {\it comparison} to observed datasets. 
The task of the model is to predict the entire star 
formation history of every type of galaxy, in all environments, 
starting from a set of primordial density fluctuations. 
We also predict the size and metallicity of the galaxy disk and bulge, 
the rotation speed of the disk, the amount of cold gas, 
the amount of hot gas in the halo and the morphology.
Once we have normalised our model to the local luminosity function 
or Tully-Fisher relation there is no {\it a priori} reason to expect to 
obtain the reasonable level of agreement we find with observables, 
such as for example, the observed 
star formation history of the universe or the abundance and 
clustering of Lyman break galaxies. 
The semianalytic approach is complementary to fully numerical 
simulations with gas; several of the rule parameterisations used are 
calibrated against numerical results whilst the semianalytic models 
can explore a much wider parameter space than is feasible with numerical 
simulations.

The models have successfully reproduced and predicted a wide range of 
galaxy observables: global properties including the shape of the luminosity 
function,  colours, faint counts, redshift distributions (Lacey et al 1993, 
Kauffmann et al 1993, 1994, Cole et al 1994, Heyl et al 1995) and more specific 
effects such as the growth of brightest cluster galaxies (Arag\'{o}n-Salamanca, 
Baugh \& Kauffmann 1998).
In this review we restrict our attention to two areas that have a 
direct bearing on the subject of this meeting - the formation of 
elliptical galaxies and the properties of galaxies at high redshift.

\section{The formation of elliptical galaxies}

\begin{figure}
{\epsfxsize=12.truecm \epsfysize=11.truecm 
\epsfbox[-100 0 580 720]{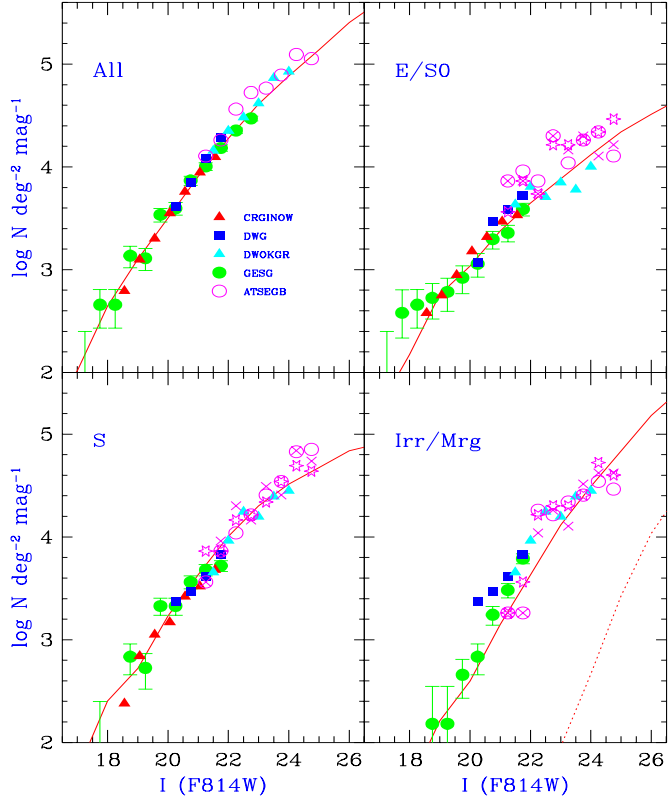}}
\caption{Faint galaxy counts from the Hubble Space Telescope, 
separated by morphological type. 
The points show the observations - full references are given in 
Baugh, Cole \& Frenk 1996b. The lines shows the predictions 
of the semianalytic models, in which the morphological type is 
assigned according to the bulge to total luminosity ratio in the I band.
The dotted line in the bottom right panel shows the contribution to the 
irregular/peculiar class  of galaxies that 
have experienced a recent merger  -- the remainder 
of this class in our model is made up of galaxies with very small bulges.
} 
\label{fig:counts}
\end{figure}

The traditional picture of elliptical galaxy formation as 
a monolithic collapse involving a single burst of star 
formation at high redshift (Eggen, Lynden-Bell \& Sandage 1962) has been 
challenged by two recent observations. 
Kauffmann, Charlot \& White (1996) have shown, using data 
from the CFRS (Lilly et al 1995) and from the Hawaii Deep 
Survey (Cowie et al 1996) that only 
one third the number of nearby bright elliptical galaxies seen today 
were already in place at $z \sim 1$ or had the colours of old, 
passively evolving stellar populations.  
In an analysis of deep optical and infrared images, Zepf (1997) has 
demonstrated that too few galaxies with very red colours are seen when  
compared with the number expected if ellipticals formed exclusively 
at high redshift. 

Semianalytic models propose a scheme whereby galactic bulges 
result from galaxy mergers (Kauffmann et al 1993, 
Kauffmann 1995, 1996; Baugh, Cole \& Frenk 1996a), 
as originally suggested by Toomre (1977).
In the models, quiescent star formation builds a disk, whilst 
material accreted during mergers is added to a bulge component. 
A major merger, in which the primary galaxy accretes 
more than some specified fraction of its own mass, (typically 
$30 \%$ or more), results in the destruction of the stellar disk 
and a burst of star formation, with all stars being placed in the 
bulge. 
Hence in a merger picture, the morphology of a galaxy changes with time.
After a major merger, a galaxy will initially be a pure bulge system and then 
quiescent star formation can start to form a new disk.

Such models have been able to reproduce a range of observations that 
distinguish between galaxies according to their morphologies. 
Baugh, Cole \& Frenk 1996b demonstrated that semianalytic models 
could explain the form of the faint counts of galaxies from HST images separated  
by morphological type (full references are given 
in Baugh et al) - Figure \ref{fig:counts}.  
Kauffmann 1995 and Baugh, Cole \& Frenk 1996a have shown that 
the population of galaxies found in model clusters at different redshifts 
exhibit evolution of the form detected by Butcher \& Oemler (1984).

A key observation in pinning down the star formation history of 
spheroidal systems is the colour-magnitude relation for 
cluster E and SO galaxies (Bower, Lucey \& Ellis 1992). 
Perhaps counter-intuitively, semianalytic models naturally reproduce 
the small scatter observed in the colours of early type galaxies 
in clusters (Kauffmann 1996, Baugh, Cole \& Frenk 1996a) and also the 
slope of the colour-magnitude relation when chemical enrichment 
is incorporated (Kauffmann \& Charlot 1997).

Baugh, Cole \& Frenk (1996a) found that $50\%$ of bright ellipticals 
taken from all environments in their model experience a major merger 
between $z=0$ and $z=0.5$. 
At these redshifts, typically only around $5\%$ of the mass of the final 
galaxy is formed in the burst of star formation that accompanies 
the major merger. The bulk of the stars have already formed in the  
progenitors that merge together -- the major merger represents the 
assembly of these stars into an elliptical. This can be contrasted with 
the situation at high redshift; an elliptical that experiences a major merger 
between $z=1.5$ and $z=3$ can form around $30\%$ of its final stellar 
mass in the accompanying burst.
Kauffmann (1996) makes the same point by showing that the mean age of stars in 
cluster ellipticals is more than $10$ Gyr, whilst the last major  merger 
occured on average $7$ Gyr ago.

\section{The star formation history of the universe}

\begin{figure}
{\epsfxsize=11.truecm \epsfysize=9.truecm 
\epsfbox[-120 100 580 700]{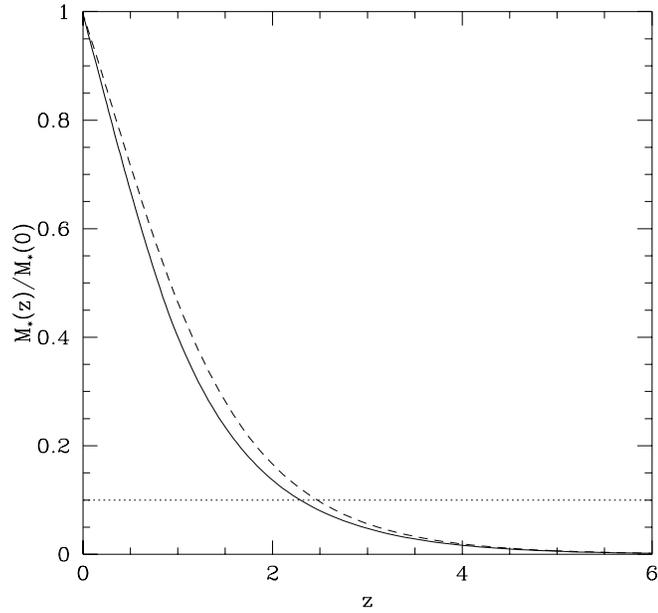}}
\caption{
The global build up of stellar mass in two hierarchical structure formation 
models. The curves show the fraction of the present day mass in stars that was in  
place by a given redshift. The solid line shows standard CDM and the dashed line 
shows a flat, low density CDM model. 
The horizontal line marks $10\%$ of the present day stellar mass. The close 
agreement between the two cosmologies can be traced to the strong feedback used and 
the normalisation of the density fluctuations to reproduce the present day abundance of 
rich clusters.
} 
\label{fig:mstars}
\end{figure}

\begin{figure}
\begin{picture}(300, 300)
\put(0, 150)
{\epsfxsize=11.truecm \epsfysize=6.8truecm 
\epsfbox[40 400 580 740]{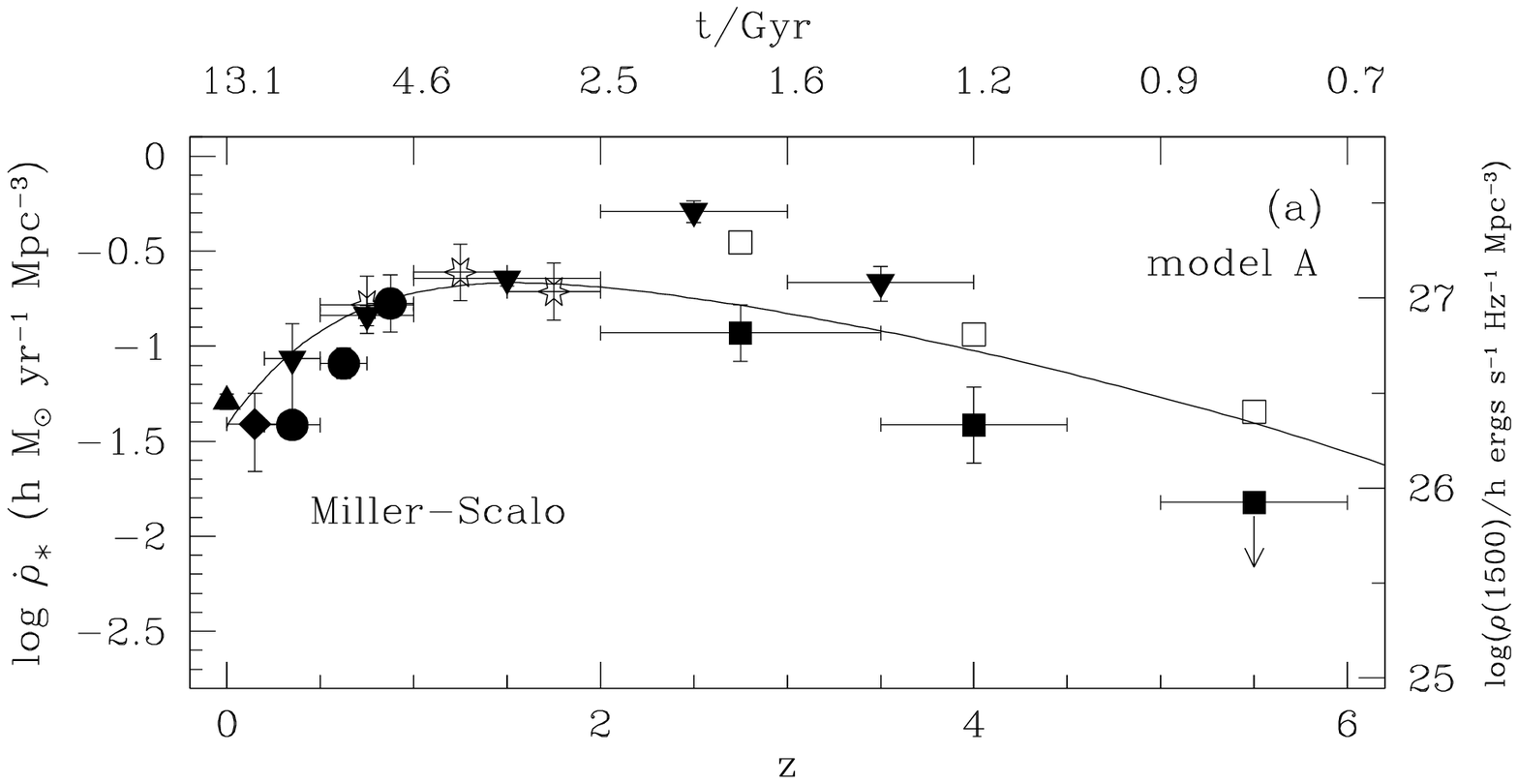}}
\put(0, -20)
{\epsfxsize=11.truecm \epsfysize=6.8truecm 
\epsfbox[40 400 580 740]{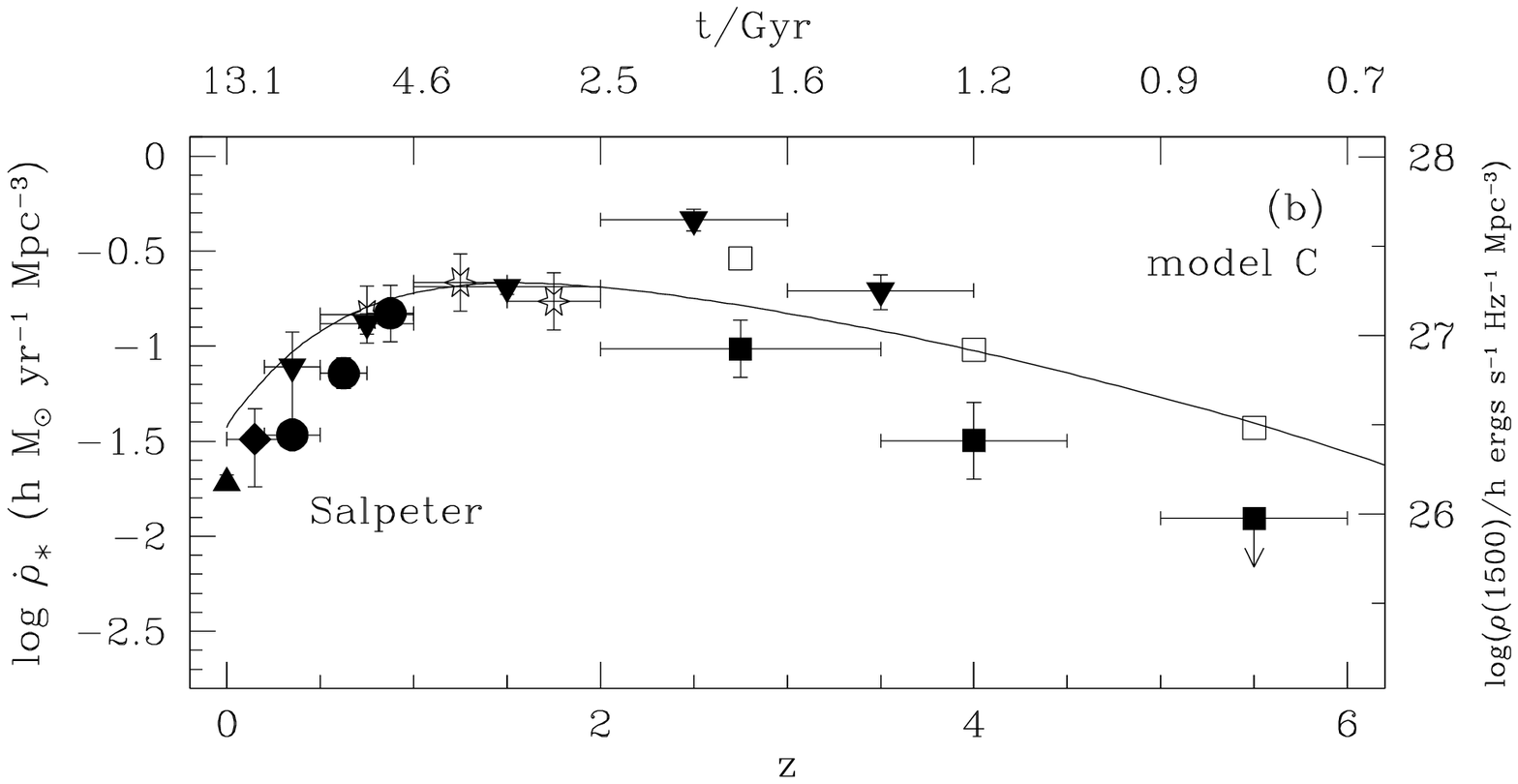}}
\end{picture}
\caption{The cosmic star formation history taken from 
Baugh et al (1998). The lines show 
the predictions of the semianalytic models 
for the star formation rate per comoving 
volume in a CDM universe 
with the critical density, which is essentially the prediction 
made in Cole et al (1994).
The right hand axis shows the corresponding flux density at $1500 {\rm \AA}$.
The symbols show a number of determinations of the star formation rate 
at different redshifts, using the luminosities at different rest 
frame wavelengths. The data are taken from the following references: 
triangle - Gallego et al (1995); diamond - Treyer et al (1997); circles - 
Lilly et al (1996) ; inverted triangles - Sawicki et al (1997); stars - 
Connolly et al (1997); filled squares - Madau et al (1996). 
The conversion from UV flux to star formation rate depends upon the IMF:  
(a) shows a Miller-Scalo IMF and (b) shows a Salpeter IMF. The open squares 
show a correction of a factor of 3 (Pettini et al 1997) to the Madau et al 
points to account for possible obscuration by dust.} 
\label{fig:madau}
\end{figure}

Hierarchical clustering theories naturally predict that galaxy formation 
occured recently, reflecting the way in which dark matter halos are assembled.
Cole et al (1994) showed that typically $50\%$ of the stellar mass at $z=0$ has 
formed since $z=1$ in certain CDM models (Figure \ref{fig:mstars}) -- this is 
due to a combination of the strong feedback employed and the normalisation 
of the density fluctuations to reproduce the abundance of rich clusters. 
At $z=3$ a mere $5\%$ of today's global stellar mass was in place. 
The observed star formation history of the universe (Madau et al 1996), 
plotted as the time derivative of Figure \ref{fig:mstars} is remarkably close 
to the predictions of the semianalytic model (Baugh, Cole, Frenk \& 
Lacey 1998) (Figure \ref{fig:madau}).
Again, it is important to stress that the model parameters are set to produce 
a reasonable match to the local galaxy luminosity function; 
indeed none of the data points 
in Figure \ref{fig:madau} existed when Cole et al was published.
The cosmic star formation history is a genuine prediction of the model -- 
none of the model parameters have been `tuned' to give a `good fit' to the 
observed star formation rate density. 

The interpretation of the data points in Figure \ref{fig:madau} is 
subject to a number of caveats. First, 
the conversion from $H\alpha$ or UV flux to star formation rate depends 
upon the form of IMF adopted; for example the amount of 
flux at $1500 {\rm \AA}$ produced by a star formation rate of one 
solar mass per year is three times higher 
with a Miller-Scalo IMF than if a Scalo IMF is used -- both 
these IMFs are compatible with local determinations of the form of the 
IMF. 
Second, the observations generally probe a limited range of galactic 
star formation rates. To get the integrated rate per unit volume, 
some form of extrapolation is necessary and is usually done by 
fitting a Schechter function to the observed star formation 
rates.
Last, and perhaps most uncertain, is the correction for the presence 
of dust in the primeval galaxies. Even a small amount of dust will attenuate 
the UV flux, leading to a potentially serious underestimate of the amount 
of star formation in the galaxy. 
Upper limits (Kashlinsky et al 1996) and tentative detections (Puget 
et al 1996) of the infrared background light currently provide some  
constraints on the amount of starlight 
from galaxies that can be reprocessed into the 
infra-red by dust (Madau, Pozzetti \& Dickinson 1997). 
A comparison of the intrinsic colours of high redshift galaxies with 
the observed, dust reddened colours is possible,  although this is sensitive to 
the choice of model galaxy and to the form of the dust extinction law adopted.
For the most extreme case in which the model galaxies are young 
starbursts, and therefore extremely blue, and the greyest dust extinction curve
is used, a correction to the inferred star formation rate by a factor of up to 
ten is suggested (Meurer et al 1996). 
For less extreme assumptions a factor in the range $1.5-3$ is advocated 
(Pettini et al 1997, Dickinson et al in preparation).

\section{The clustering of Lyman-break galaxies}

The extraction of high redshift galaxies from deep images 
of the sky using exposures taken in several filters has made the 
detection of large numbers of primeval galaxies possible 
(Steidel \& Hamilton 1992).
High redshift star forming galaxies are identified by their Lyman break moving 
through one of the filters, giving a red colour, whilst the other 
filters indicate a blue colour.
This technique has been applied from the ground (Steidel et al 1996) and 
with the HST (Madau et al 1996) to identify galaxies in the redshift range 
$z=2-4.5$ (Steidel et al 1997; Lowenthal et al 1997).

Using the full colour selection employed by Steidel et al (1996) and including  
attenuation of the light due to intervening cold gas (Madau 1995), 
semianalytic models have demonstrated that a range of CDM models can reproduce 
the observed abundance of Lyman break galaxies, in spite of  
the fact that typically only 
$5\%$ of the stars that will have formed by today are already in place 
at this time (Baugh et al 1998). 
The models give the mass of the dark matter halo that hosts the Lyman break 
galaxy, allowing a bias parameter for these objects to be computed using the 
formalism developed by Mo \& White (1996). Baugh et al find a bias between 
fluctuations in Lyman break galaxies and the fluctuations in the underlying 
density distribution of $b=4$ at $z=3$. Since the Lyman break galaxies are found 
to form in the most massive halos that have collapsed at high redshift, it is 
natural for these objects to be highly biased tracers of the mass distribution.

\begin{figure}
{\epsfxsize=12.truecm \epsfysize=8.truecm 
\epsfbox[90 230 580 610]{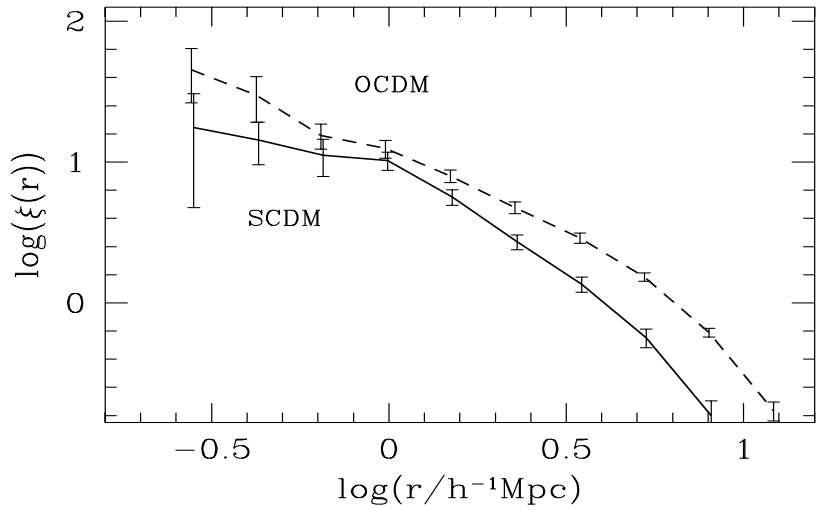}}
\caption{
The correlation function of Lyman break galaxies, computed by taking 
the halos identified in a high resolution N-body simulation at $z=3$ and 
using the semianalytic model of galaxy formation to predict which halos 
should contain Lyman break galaxies (Governato et al 1998). 
Two cosmologies are shown standard CDM (SCDM) and a low density open model 
(OCDM). The errorbars are bootstrap estimates.
} 
\label{fig:xi}
\end{figure}

The observation of large numbers of Lyman break galaxies, followed up by 
spectroscopic confirmation of their redshifts, will soon make it possible 
to measure the angular and spatial correlation functions of these objects.
Baugh et al (1998) use their computation of the bias parameter of the Lyman 
break galaxies to make an estimate of these correlation functions.
The approximations used in this calculation break down on small scales where 
the clustering signal is strongest.
 
In order to improve the accuracy of these predictions,  
Governato et al (1998) combined semianalytic modelling with 
high resolution N-body simulations. 
Dark matter halos are identified in the simulation 
at $z=3$ and the semianalytic galaxy formation model is run 
for each halo mass. 
Halos that the semianalytic model predicts should contain a Lyman break  
galaxy are labelled and the correlation function of these objects is 
measured (Figure \ref{fig:xi}). 
Some of the more massive dark matter halos contain more than one 
Lyman break galaxy. 
The bias parameters measured in the simulation are in good agreement with those 
predicted analytically.
Steidel et al (1997) have discovered a large concentration of Lyman break 
galaxies in one of their fields (see also the contribution of John 
Peacock to this volume). Using our simulations we have found that such 
structures are not unexpected, even in a standard Cold Dark Matter simulation. 
The semianalytic model allows us to reach this conclusion without having 
to resort to making uncertain assumptions about 
the masses of halos that contain Lyman break galaxies or about the number of 
these objects per halo -- indeed the result that a dark matter halo can contain 
more than one Lyman break galaxy has an important bearing on the assessment of 
the significance of the observed concentration of Lyman break galaxies.

\section{Conclusions}

This is an exciting period for the study of galaxy formation which promises 
to continue with the construction of more large telescopes and 
observations of galaxies being carried out at many different redshifts. 
Many of the details of the galaxy formation process remain unknown and 
are currently inaccessible to numerical investigation.
However, in view of the remarkable successes enjoyed by semianalytic models, 
especially when one considers the magnitude of the task attempted, 
it would appear that any future, more complete theory of galaxy formation 
will share many features in common with the models discussed here.

\begin{acknow}
This research  was supported 
by the European Commission through the TMR Network on `` The Formation 
and Evolution of Galaxies'' and in part by a PPARC rolling grant.
SMC acknowledges a PPARC Advanced Fellowship and CSF
acknowledges a PPARC Senior Fellowship. CGL was supported by the Danish
National Research Foundation through its establishment of the Theoretical
Astrophysics Center.
\end{acknow}

\end{document}